\documentclass[referee,sn-mathphys-num]{sn-jnl}

\usepackage{float}
\usepackage{miller}
\usepackage{xcolor}
\usepackage{graphicx}

\usepackage[version=3]{mhchem} 

\newcommand{\InSe}{$\alpha$-In\textsubscript{2}Se\textsubscript{3}}

\begin{document}

\title[Article Title]{Field-effect transistors based on charged domain walls in van der Waals ferroelectric \InSe{}}


\author[1,8]{\fnm{Shahriar Muhammad} \sur{Nahid}}\email{snahid2@illinois.edu}

\author[2,8]{\fnm{Haiyue} \sur{Dong}}\email{haiyued2@illinois.edu}

\author[3,8]{\fnm{Gillian} \sur{Nolan}}\email{gnolan2@illinois.edu}


\author[3,8]{\fnm{Andre} \sur{Schleife}}\email{schleife@illinois.edu}

\author[4,5]{\fnm{SungWoo} \sur{Nam}}\email{sungwoo.nam@uci.edu}

\author[3,7,8]{\fnm{Pinshane Y.} \sur{Huang}}\email{pyhuang@illinois.edu}

\author[6]{\fnm{Nadya} \sur{Mason}}\email{nmason1@uchicago.edu}

\author*[1,3,7,8]{\fnm{Arend M.} \sur{van der Zande}}\email{arendv@illinois.edu}

\affil[1]{\orgdiv{Department of Mechanical Science and Engineering}, \orgname{University of Illinois Urban--Champaign}, \orgaddress{\city{Urbana}, \postcode{61801}, \state{IL}, \country{United States}}}

\affil[2]{\orgdiv{Department of Physics}, \orgname{University of Illinois Urbana--Champaign}, \orgaddress{\city{Urbana}, \postcode{61801}, \state{IL}, \country{United States}}}

\affil[3]{\orgdiv{Department of Materials Science and Engineering}, \orgname{University of Illinois Urbana--Champaign}, \orgaddress{\city{Urbana}, \postcode{61801}, \state{IL}, \country{United States}}}

\affil[4]{\orgdiv{Department of Mechanical and Aerospace Engineering}, \orgname{University of California, Irvine}, \orgaddress{\city{Irvine}, \postcode{92697}, \state{CA}, \country{United States}}}

\affil[5]{\orgdiv{Department of Materials Science and Engineering}, \orgname{University of California, Irvine}, \orgaddress{\city{Irvine}, \postcode{92697}, \state{CA}, \country{United States}}}

\affil[6]{\orgdiv{Pritzker School of Molecular Engineering}, \orgname{University of Chicago}, \orgaddress{\city{Chicago}, \postcode{60637}, \state{IL}, \country{United States}}}

\affil[7]{\orgdiv{Materials Research Laboratory}, \orgname{University of Illinois Urbana--Champaign}, \orgaddress{\city{Urbana}, \postcode{61801}, \state{IL}, \country{United States}}}

\affil[8]{\orgdiv{Grainger College of Engineering}, \orgname{University of Illinois Urban--Champaign}, \orgaddress{\city{Urbana}, \postcode{61801}, \state{IL}, \country{United States}}}

\abstract{

Charged domain walls (CDW) in ferroelectrics are emerging as functional interfaces with potential applications in nonvolatile memory, logic, and neuromorphic computing. However, CDWs in conventional ferroelectrics are vertical, buried, or electrically inaccessible interfaces that prevent their use in functional devices. Here, we overcome these challenges by stacking two opposite polar domains of van der Waals ferroelectric \InSe{} to generate artificial head-head (H-H) CDWs and use edge contact to fabricate charged domain wall-based field-effect transistors (CDW-FET). We relate the atomic structure to the temperature-dependent electrical and magneto-transport of the CDW-FET. CDW-FETs exhibit a metal-to-insulator transition with decreasing temperature and enhanced conductance and field-effect mobility compared to single domain \InSe{}. We identify two regimes of transport: variable range hopping due to disorder in the band edge below 70 K and thermally activated interfacial trap-assisted transport above 70 K. The CDW-FETs show room-temperature resistance down to 3.1 k$\Omega$ which is 2-9 orders of magnitude smaller than the single CDW in thin-film ferroelectrics. These results resolve longstanding challenges with high CDW resistance and their device integration, opening opportunities for gigahertz memory and neuromorphic computing.}


\maketitle

\section*{Main}

Ferroelectric domain walls are emerging as novel functional interfaces, where their reduced dimensionality and different symmetry generate distinct properties not observed in adjacent bulk domains\cite{meier2022,sharma2022,nataf2020,bednyakov2018} such as polarity,\cite{Aret2012} magnetic order,\cite{Geng2012,giraldo2021} enhanced electromechanical coupling,\cite{sluka2012} mechanical softness,\cite{Stefani2020} or conducting states.\cite{seidel2009, Zhang2019, Liu2021, Maksymovych2011, Farokhipoor2011, sluka2013, werner2017, jiang2020, mundy2017, wu2012, choi2010, mcquaid2017, oh2015, stolichnov2015, risch2022, lindgren2017} Of particular interest are charged domain walls (CDWs), where adjacent domains with opposite polarization meet to create interfacial bound charges which require screening by free carriers. As a result, CDWs display several orders of magnitude higher conductivity compared to their bulk domains.\cite{seidel2009, Farokhipoor2011, sluka2013, werner2017, jiang2020, mundy2017, wu2012, choi2010, mcquaid2017, oh2015, stolichnov2015, risch2022, lindgren2017} While the enhanced CDW conductivity was predicted nearly 50 years ago,\cite{Vul1973} it was not until 2009 that this phenomenon was first experimentally confirmed in BiFeO\textsubscript{3}.\cite{seidel2009} After this discovery, the field rapidly expanded, leading to the realization of conducting CDWs in many different ferroelectrics,\cite{meier2022, sharma2022} such as BaTiO\textsubscript{3}, PZT, LiNbO\textsubscript{3}, ErMnO\textsubscript{3}, KTiOPO\textsubscript{4}, and YMnO\textsubscript{3}. In most cases, the domain walls are oriented along the out-of-plane direction and exhibit orders of magnitude lower resistance\textemdash on the order of M$\Omega$ to T$\Omega$\textemdash compared to the surrounding bulk domains, as revealed by conducting atomic force microscopy or vertical transport characterization techniques.\cite{seidel2009, mundy2017, jiang2020, sluka2013} Other systems show mobile CDWs under out-of-plane electric field, but these CDWs cannot be electrically accessed for transport because they are buried inside the domains.\cite{liu2023} These exciting results have led to potential applications of these reconfigurable CDWs in nonvolatile memory, logic, and artificial synapses.\cite{meier2022, sharma2022, Jiang2019} These applications demand new strategies to create in-plane, electrically accessible, and gate-tunable CDWs that may be integrated into devices as well as a more complete understanding of the mechanisms and limits of interfacial conduction. 

However, such investigations have been challenging due to several limiting factors in conventional ferroelectrics. First, CDWs in these ferroelectrics are out-of-plane interfaces,\cite{seidel2009, sluka2013, werner2017} allowing only two-point transport measurements, or are buried inside the bulk ferroelectric,\cite{liu2023} making them electrically inaccessible. These limitations prohibit the implementation of standard characterization techniques. Second, the controlled and reliable generation of single CDWs remains a challenge.\cite{sharma2022} Finally, the CDW resistance in conventional ferroelectrics, as measured using two-point transport methods, is between M$\Omega$ to T$\Omega$ range, making it impossible to drive read-out circuits at GHz speed due to the Johnson-Nyquist limit.\cite{Nyquist1928,Johnson1928}

In contrast to conventional thin-film ferroelectrics, van der Waals ferroelectrics show fundamentally new phenomena arising from their layered and anisotropic structure that open new opportunities for CDW-based devices. For example, they exhibit stable polarization in the monolayer limit,\cite{Xue2018} integrability with dissimilar materials,\cite{zhang2023} quadruple well potential,\cite{Brehm2020} metallic ferroelectricity,\cite{fei2018} bending angle dependent polarization reversal,\cite{Han_SMN_2023} or polarity from nonpolar parent materials;\cite{Tsymbal2021} none of which exists in conventional ferroelectrics. Furthermore, the absence of surface dangling bonds facilitates heterostructure formation and controlled generation of otherwise unstable interfaces. This ability allows for the artificial stacking of opposite polar domains, thereby enabling the formation of CDWs with electrical accessibility.

\begin{figure*}[ht]
    \centering
    \includegraphics[width=131 mm]{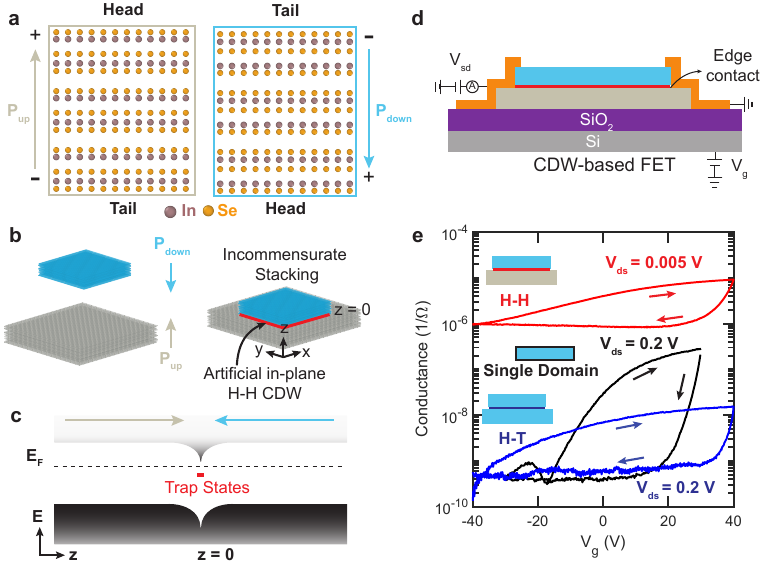}
    \caption{\textbf{Fabrication of H-H CDWs, their band diagram, and integration in CDW-based field effect transistors (CDW-FETs.} \textbf{a,} Atomic structure of \InSe{} showing up and down polarization. The positive and negative sides are denoted as Head and Tail. \textbf{b,} Schematic of an in-plane artificial H-H CDW formed by incommensurately stacking two oppositely polarized flakes ($P_{down}$, denoted in cyan, and $P_{up}$, denoted in tan) of van der Waals ferroelectric \InSe{}. \textbf{c,} Out-of-plane band diagram showing band bending at the H-H CDW.  \textbf{d,} Schematic of a CDW-FET using edge contact, \textbf{e,} Comparison of the conductance versus gate bias curves of a CDW-FET with single domain \InSe{} and Head-Tail heterostructure-based FETs at room temperature. The CDW-FET shows enhanced conductance for all gate bias compared to the other structures.}
    \label{fig:1}
\end{figure*} 

In this work, we stack two oppositely polarized \InSe{} flakes to generate artificial in-plane H-H CDWs and use them to fabricate CDW-based field-effect transistors (CDW-FETs). These CDW-FETs show conductance approximately 4 orders of magnitude higher than that of the surrounding bulk domains. Remarkably, they exhibit the highest conductance ever reported for a single CDW, exceeding those in conventional ferroelectrics by more than 2 to 9 orders of magnitude. We use aberration-corrected electron microscopy to explore the atomic structure of the interface and temperature-dependent transport to probe the limits of conductivity. This work provides new insights into the transport mechanisms and role of disorder in CDW conduction and enables a new class of device structures based on ferroelectric CDWs that are not achievable in conventional thin-film ferroelectrics.

Figure \ref{fig:1} outlines the concept of creating CDW-FETs from \InSe{} heterostructures. As shown in Figure \ref{fig:1}a, \InSe{} is a van der Waals ferroelectric, where the distortion of the central Selenium atomic plane within each layer leads to out-of-plane polarization.\cite{Michael2018} Similar to all dielectrics, polarization is defined as a vector, pointing from negative (tail) to positive (head) charge. As shown in Figure \ref{fig:1}b, stacking two multilayer \InSe{} flakes, where the bottom flake has up polarization and the top flake has down polarization, leads to the formation of incommensurate H-H CDWs. See \nameref{Methods} and Supplementary Section 1 for details on the heterostructure fabrication. Piezoelectric force microscopy (PFM) confirms the relative polarization of each flake (Supporting Fig. 1). Supplementary Section 2 discusses the various possible heterostructures formed by this process and evaluates how frequently each occurs. We only choose H-H CDWs for subsequent CDW-FET fabrication. Figure \ref{fig:1}c shows the hypothesized out-of-plane (z-direction) band diagram of the H-H CDW, which governs the in-plane (x-y direction) transport behavior. Since \InSe{} is a semiconductor, the local electrostatic potential from the positive bound charges results in a downward band-bending at the CDW, bringing the conduction band minima closer to the Fermi energy ($E_F$). The localized band bending at the interface leads to free electrons and an in-plane conducting state. Additionally, any disorder at the interface will lead to trap states, shown as the red block.

Figure \ref{fig:1}d shows the schematic of a CDW-FET, fabricated from these H-H CDWs (see \nameref{Methods} and Supplementary Section 1). We use electron-beam (ebeam) lithography to add electrical contacts to the heterostructure. Electrical access to the interfacial CDW is achieved through edge contact, where the electrodes drape across the step of the top flake, similar to techniques employed in two-dimensional material heterostructures.\cite{Choi2022} Supplementary Fig. 2 shows an STEM image of the edge contact. The SiO\textsubscript{2}/Si substrate serves as the global back gate.


Figure \ref{fig:1}e compares the conductance versus gate bias transfer curves of a CDW-FET (labeled H-H \#1) with head-tail (H-T) heterostructure and single-domain flake-based FETs, including both up and down sweeps. Supplementary Table 1 summarizes the dimensions of each device. The measurement is at 300 K.

\begin{figure*}[ht]
    \centering
    \includegraphics[width=161.45 mm]{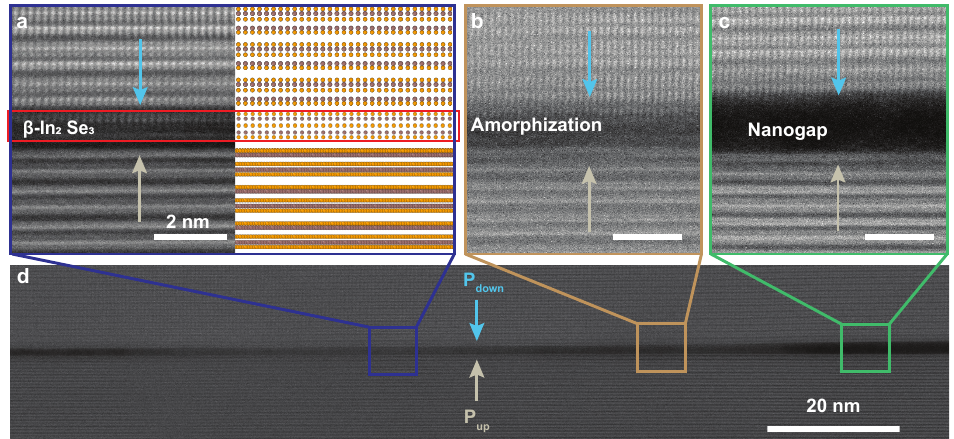}
    \caption{\textbf{Atomic structure of H-H CDW in \InSe{}}. \textbf{b-d,} Cross-sectional STEM images of the atomic structure of the H-H CDW show uniform domains in the top and bottom flake. However, the CDW exhibits interfacial inhomogeneity including (b) the formation of $\beta-In_2Se_3$ phase (red rectangle), (c) amorphization of \InSe{}, and (d) nanogap. \textbf{e,} Low-magnification image of the H-H CDW layer showing the relative location of each region.}
    \label{fig:2}
\end{figure*}

Both the FETs bassed on single domain \InSe{} and H-T heterostructure display n-type semiconductor transport behavior with large hysteresis. The single domain transport is consistent with previous literature.\cite{si2019} In contrast, the CDW-FET has a drastically higher conductance at all gate biases\textemdash around 30 and 600 times higher than the single domain and H-T, respectively, at 30 V gate bias. Moreover, the CDW-FET channel appears heavily doped with less than 1 order of magnitude modulation in conductance under gating.  Supplementary Figs. 3-4 show the corresponding output and gate/drain current versus gate bias curves and confirm that the CDW-FET have ohmic contacts with negligible gate leakage. Importantly, the H-T heterostructure, despite containing a stacked interface, does not show enhanced conductance, indicating that the presence of a stacked interface alone is insufficient to account for the high conductance observed in the CDW-FET. This observation supports the hypothesis that the enhanced conductance and the absence of an off-state arise from band bending effects unique to the H-H configuration, as illustrated in Figure \ref{fig:1}c. Additionally, both the CDW-FET and H-T FET show larger hysteresis than the single domain, indicating potential interfacial trap states independent of the interface type. We also check the robustness of the conductance enhancement. We fabricated three separate CDW-FETs, labeled H-H \#1-3. All the results in Figure \ref{fig:2}-\ref{fig:5} are from H-H \#1. H-H \#2-3 show qualitatively similar results, with transport data shown in Supplementary Figs. 5 and 6.


In Figures \ref{fig:2}-\ref{fig:5}, we relate the atomic structure with the temperature, gate voltage, and magnetic field dependent electrical transport of the same CDW-FET (H-H \#1) to understand how the structure of the van der Waals interface in the H-H CDW defines the transport mechanisms and limits of performance. We note that the in-plane structure and presence of the electrostatic gate allow control of the free carrier density, providing insights into the CDW conduction mechanisms that have previously not been possible with out-of-plane thin-film CDWs, where the free carrier density is typically unknown.

First, in Figure \ref{fig:2}, we examine the interfacial structure of the CDW-FET (H-H \#1). After completion of all transport measurements, we use focused ion beam to cross-section the heterostructure, then image the interface using aberration-corrected annular dark-field scanning transmission electron microscopy (ADF-STEM) (see \nameref{Methods}). Figure \ref{fig:2}a-c show cross-sectional STEM images of different regions of the H-H CDW, while Figure \ref{fig:2}d is the low-magnification image showing the relative location of each region. Figure \ref{fig:2}a shows both the STEM image of a single region in the CDW and the corresponding schematic of the crystal structure. The position of the central selenium atomic plane in each layer confirms that the top flake has down polarization, whereas the bottom flake has up polarization, creating a H-H CDW at the interface marked by the red block. The polarization of each flake is the same across all the imaged regions, and is in agreement with the PFM images in Supplementary Fig. 1. Note that incommensurate stacking results in a misalignment between the top and bottom flakes. The top flake is imaged along the major axis, while the atoms of the bottom flakes are off-axis. Thus, the individual atoms are resolved in the top flake, while only the atomic planes are resolved in the bottom flake.

There are three important observations rising from the interface structure shown in Figure \ref{fig:2}a-c. First, there is reconstruction at the H-H interface, which accommodates the interfacial polar discontinuity and the depolarization field. Similar reconstruction commonly occurs in conventional thin-film ferroelectric CDWs.\cite{jia2008,liu2023} Second, the out-of-plane width of the CDWs ranges between 1-3 nm, and remains continuous over micrometer length scales. In contrast, CDWs in thin-film ferroelectrics often shift between different atomic planes, resulting in discontinuous or meandering interfaces.\cite{Gonnissen2016, jia2008} The CDW continuity is critical for preserving its conductive properties and reliable integration into functional electronic devices. 

Third, the structure at the interface is heterogeneous. For example, in the different regions shown in Fig. \ref{fig:2}a-c, the interface consists of a monolayer of the centrosymmetric $\beta-In_2Se_3$ phase (Fig. \ref{fig:2}a), amorphous material (Fig. \ref{fig:2}b), or a nanogap (Fig. \ref{fig:2}c). Supplementary Fig. 7 shows additional structures that also exist in the same interface, which include steps and partial phase changes, and Supplementary Fig. 8 shows electron energy loss spectroscopy analysis of the nanogap and amorphous regions. As shown in the lower magnification image in Fig. \ref{fig:2}d, the different structures all coexist in close proximity with in-plane separation of 20-90 nm. The $\beta-In_2Se_3$ structure is consistent with the natural H-H CDWs that sometimes occur in exfoliated \InSe{} (see Supplementary Fig. 9), and is predicted to have new bands compared to the bulk \InSe{}.\cite{nolan2024,wu2024} From the EELS analysis in Supplementary Fig. 8, the amorphous interface does not contain significant amounts of oxygen or carbon and does not have a significant drop in Indium concentration. Moreover, the width of the amorphous region does not change significantly from the size of a \InSe{} monolayer. Thus, we hypothesize that the amorphous region is reconstructed \InSe{} with domains smaller than the thickness of the cross-section (~ 15 nm). Meanwhile, the nanogap contains residual carbon and typically exists near step edges. Thus, the significant number of steps in exfoliated \InSe{} flakes leads to regions of delamination due to the bending dynamics of 2D materials.\cite{Han2020, Han_SMN_2023} The residual carbon is either trapped during transfer, or permeates during FIB etching. This heterogeneity in the interface and width of the CDW will lead to different electronic states or density of bound charges across the interface.
 
\begin{figure*}[ht]
    \centering
    \includegraphics[width=118 mm]{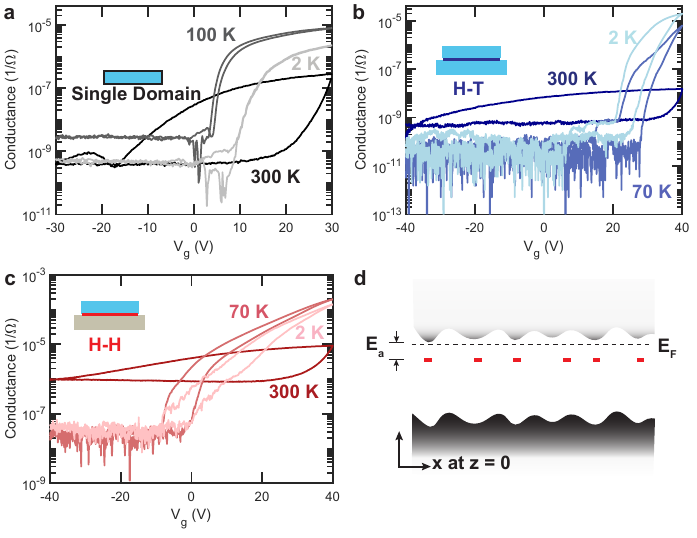}
    \caption{\textbf{Temperature dependent electrical transport of CDW-FET and the corresponding band diagram.} \textbf{a-c,} Conductance versus gate bias transfer curves of single domain, H-T heterostructure, and CDW-FET at different temperatures. Both single domain and H-T heterostructure show n-type transport at all temperatures. In contrast, the CDW-FET is "always on" with respect to gate bias at 300 K, but transitions to n-type transport at lower temperatures. The hysteresis in single domain \InSe{} reduces significantly at 2 K, while both the H-T and CDW-FET maintains moderate hysteresis throughout the temperature range. \textbf{d,} The corresponding in-plane band diagram of CDW-FET. Interfacial heterogeneity results in in-plane modulation of the conduction band minima and generates spatially varying trap states (marked as red block) with an activation energy $E_a$. At high temperatures, the trap states activate leading to "always on" transport in CDW-FET with enhanced hysteresis.}
    \label{fig:3}
\end{figure*} 

Next, we investigate how heterogeneity at the interface impacts transport. Fig. \ref{fig:3} compares the temperature-dependent transfer curves of FETs from (a) single domain \InSe{}, (b) H-T heterostructure, and (c) H-H CDW. With decreasing temperature, the single domain continues to show n-type transport, while the CDW-FET shows a gate-dependent metal-insulator transition, with increasing conductivity at positive gate bias and decreasing conductivity at negative gate bias, leading to a transition from semimetallic to n-type semiconductor behavior. Consistently, the CDW-FET maintains a much higher conductance than the H-T and single domain at all temperatures. All structures show significant hysteresis at room temperature. However, at low temperature, the hysteresis in the single domain nearly vanishes but the H-T and CDW-FET exhibit reduced but persistent hysteresis.

Figure \ref{fig:3}d shows a proposed in-plane band diagram at the CDW interface to explain the transport. We interpret the temperature-dependent transport as a combination of two mechanisms, both of which are commonly observed in two-dimensional materials under disorder like MoS\textsubscript{2}\cite{cui2015, Park2016, kim2012, jariwala2013,radisavljevic2013} and graphene.\cite{chen2008} First, the heterogeneity in width at the interface will lead to in-plane modulations of the conduction band minimum compared with the fermi energy. Second, the localized states from different phases will lead to thermally activated trap states (shown as red blocks) with an activation energy $E_a$. 

Mott variable range hopping (VRH) model describes how band edge disorder leads to electron hopping between regions of lower conduction band minima energy at low temperature:\cite{Mott1968, Mott1975} 

\begin{equation}
G= G_{0} \exp[-(\frac{T_0}{T})^{1/3}]
\label{Eq:VRH}
\end{equation}

where, $G$ is the conductance, $G_0$ is the prefactor, $T_0$ is the characteristic temperature, $T$ is the temperature.

Meanwhile, thermally activated traps lead to hysteresis in transfer curves.\cite{Park2016} The trap states get ionized above a critical temperature defined by the activation energy $E_a$. Trap states lead to an Arrhenius scaling of the hysteresis above the critical temperature:

\begin{equation}
\Delta n= n_{fix} + n_{max} \exp(-\frac{E_a}{K_BT})
\label{Eq:Arrhenius}
\end{equation}

Here, $n_{fix}$ denotes the density of temperature-independent traps (traps with an activation energy that is smaller compared to the measured temperatures), $n_{max}$ is the maximum density of thermally activated trap states, and $K_{B}T$ is the thermal energy. 

See Supplementary Section 3 for a detailed discussion of VRH and thermally activated trap-assisted transport.


\begin{figure*}[ht]
    \centering
    \includegraphics[width=173 mm]{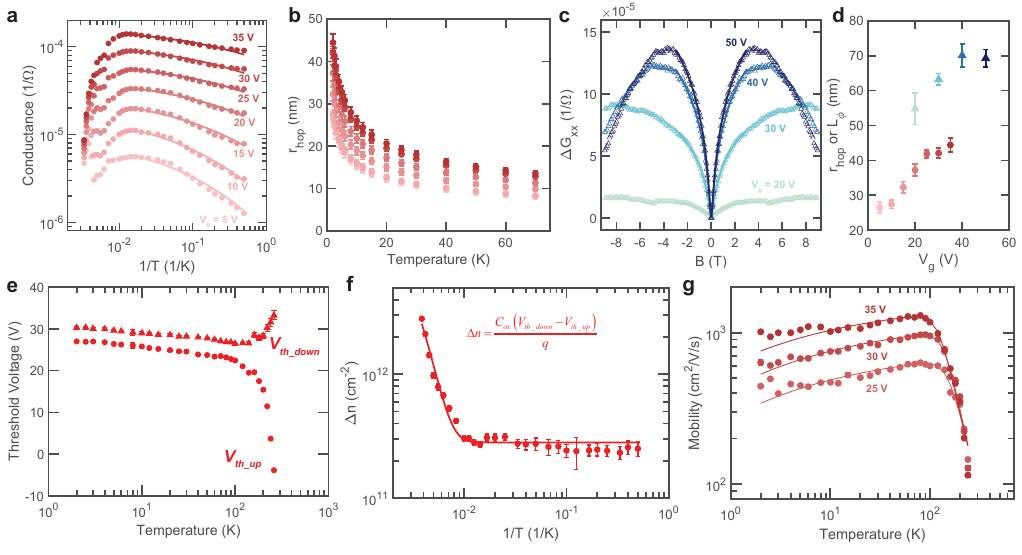}
    \caption{\textbf{Temperature scaling of various transport parameters and magneto-transport measurements for the H-H CDW confirm VRH transport below 70 K and trap-assisted transport above 70 K.} \textbf{a,} Conductance versus inverse temperature ($1/T$) for 5-35 V gate bias. For temperature $<$ 70 K, the conductance follows the scaling behavior predicted by VRH model, as indicated by the solid lines. \textbf{b,} Extracted hopping length ($r_{hop}$) with temperature for different gate biases. Inset plots $r_{hop}$ with $1/T^{1/3}$.  \textbf{c,} Change in magneto-conductance ($\Delta G_{xx}$) of H-H \#1 with magnetic field ($B$) for different gate biases at 2 K. The curves show an increase in $\Delta G_{xx}$ with $B = 3$ T, confirming the weak localization. The solid lines show the fit of the data to HLN model. \textbf{d,} Phase coherence length ($L_\phi$), extracted from the fits in (c), and $r_{hop}$, extracted from the fits in (a), increase with gate bias at 2 K. The values are on the same order as the disorders observed in the STEM images.\textbf{e,} Variation of $V_{th\_up}$ and $V_{th\_down}$, extracted from the temperature dependent transfer curves of CDW-FET, at different temperatures. The threshold voltages show a constant 3 V difference up to 70 K, and then increases with increasing temperature. \textbf{b,} Variation of trap density, calculated from the threshold voltage difference, with inverse temperature. Trap density follows thermal activation governed by Arrhenius relation. \textbf{g,} Field-effect mobility at $V_g = 25-35$ V for different temperatures. The mobility are extracted from the up-sweep branch of the transfer curves. Mobility for CDW-FET first increases with increasing temperature and then sharply drops above 70 K, showing a transition from VRH transport to thermally activated trap-assisted transport. The solid line shows the fit, which considers the two scattering mechanisms according to the Matthiessen's rule.}
    \label{fig:4}
\end{figure*} 

In Figure \ref{fig:4}, we analyze the temperature-dependent transport, low temperature magneto transport, and temperature dependent hysteresis in terms of the VRH and trap-assisted transport models. Figure \ref{fig:4}a is a logarithmic plot of conductance of CDW-FET \#1 versus $1/T$ at 5-35 V gate biases, with drain bias $V_{D} = 5$ mV. At all gate biases, the conductance increases with increasing temperature, until a transition temperature of 70 K. The transition suggests that there are different mechanisms dominating the low temperature and high temperature transport.

The low temperature scaling of the conductance is consistent with the transport in heterogeneous potential landscapes according to the VRH model. We fit the conductance to the VRH model up to 70 K, shown as the solid lines in Figure \ref{fig:4}a. Figure \ref{fig:4}b shows the corresponding average hopping distance $r_{hop}$, extracted from the fits, as a function of temperature. See Supplementary Section 3 for the details of VRH model.

As an independent probe of disorder, we also perform magnetotransport at various gate biases at 2 K. Figure \ref{fig:4}c shows the longitudinal conductance change $\Delta G{xx}$ versus magnetic field ($B$) of H-H \#1 for 20-50 V gate biases. The data show a symmetric modulation in conductivity with magnetic field, but does not show any Landau oscillations up to 9 T. The increase in $\Delta G{xx}$ with magnetic field confirms the weak localization in this system. We fit the magnetotransport data using the Hikami–Larkin–Nagaoka (HLN) model\cite{bergmann1984} up to 3 T, as shown by the solid lines in Fig. \ref{fig:4}c. The fits provide the phase coherence length $L_{\phi}$ of electrons in the system. Figure \ref{fig:4}d compares $r_{hop}$  (red tone) with $L_{\phi}$ (blue tone) at different gate biases at T = 2 K. Both parameters are related to the length scale of the disorder in the system, but are not identical. They increase with increasing gate, as an increase in Fermi energy leads to occupied states at higher energy. We note that the values of $r_{hop}$ and$L_{phi}$ are similar to the length scale at which we observe the disorders in Figure \ref{fig:2}.

Next, we assess the hysteresis in the transfer curves. Fig. \ref{fig:4}e plots the extrapolated threshold voltage of the CDW-FET transfer curve  with respect to temperature for the up ($V_{th\_up}$) and down ($V_{th\_down}$) sweeps. The threshold voltages are calculated using the linear extrapolation method.\cite{ORTIZCONDE201390} Below the critical temperature of 70 K observed in Fig. \ref{fig:4}a, $V_{th\_up}$ and $V_{th\_down}$ shift together with a difference of around 3 V, and then diverge above the critical temperature. Figure \ref{fig:4}f plots the trap density ($\Delta n$) between the up and down sweeps, calculated from the difference in threshold voltages, versus the inverse temperature $T^{-1}$. The trap density shows two regimes: a constant below the critical temperature and an exponential increase above the critical temperature. Shown as the solid line fit, this behavior is consistent with thermally activated traps,\cite{Park2016} as described by the Arrhenius relation in equation \ref{Eq:Arrhenius}. Fitting the data in Fig. \ref{fig:4}f to equation (\ref{Eq:Arrhenius}), we obtain $n_{fix}$ ($2.63\pm0.44\times10^{11}$ cm\textsuperscript{-2}), $n_{max}$ ($3.94\pm1.94\times10^{13}$ cm\textsuperscript{-2}), and $E_a$ ($64.6\pm10.3$ meV).

Fig. \ref{fig:4}g plots the field-effect mobility versus temperature for CDW-FET extracted from the up-sweep for 25, 30, and 35 V gate biases. Because of the non-uniform shape of the top-flake, the current distribution is likely also non-uniform and unknown. However, to estimate the mobility, we assume a uniform current density, which means that the reported values are the lower limits. Mobility shows a nonmonotonic trend with temperature, increasing from 2 K to 70 K and then decreasing sharply above 70 K. At 35 V gate bias and 70 K, mobility reaches approximately 1300 cm\textsuperscript{2}/V/s. Supplementary Fig. 10 shows that the mobility of single domain sample is approximately 1-3 orders of magnitude smaller across all temperatures. The trend inCDW-FET mobility with temperature is in accordance with the Mott VRH and trap-assisted transport mechanisms described above. Below 70 K, mobility scales as $exp(-\frac{T_0}{T})^{1/3}$, while above 70 K, it scales inversely with trap density $exp(\frac{E_a}{K_BT})$. We fit the mobility data with these two scaling laws using Matthiessen's rule, as shown by the solid lines in Fig. \ref{fig:4}g. The fit yields $E_a$ = 62-73 meV, similar to the value obtained from the hysteresis analysis. The close values obtained in these two independent analyses confirm the trap-assisted transport at high temperature.
 
Supplementary Fig. 11 shows the complementary temperature scaling of the extracted subthreshold swing, threshold voltage difference, and maximum mobility of the CDW-FET.


\begin{figure*}[ht]
    \centering
    \includegraphics[width=130 mm]{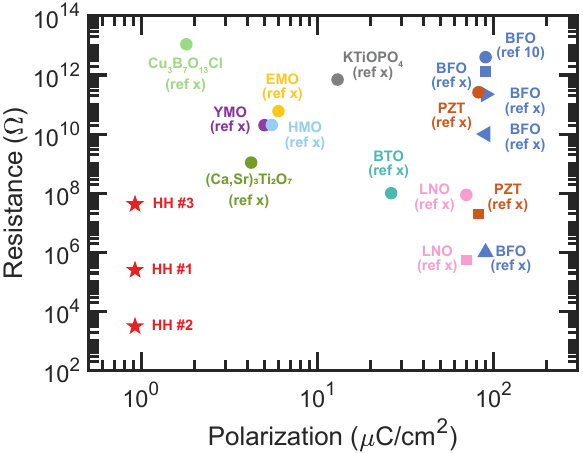}
    \caption{\textbf{Comparison of the room temperature resistance of CDW-FET in \InSe{} studied in this work with other ferroelectric domain walls.}\cite{mcquaid2017, choi2010, oh2015, mundy2017, wu2012, lindgren2017, sluka2013, jiang2020, werner2017, stolichnov2015, risch2022, Farokhipoor2011, seidel2009, zhang2023, Maksymovych2011, Liu2021} The \InSe{} H-H CDWs (H-H \#1, \#2, and \#3) are marked by red stars. H-H \#3 has the lowest resistance of 3.1 K$\Omega$, which is more than 2 to 9 orders of magnitude higher than the ones reported in the literature.}
    \label{fig:5}
\end{figure*}

Next, we put the low resistance observed in \InSe{} based CDW-FETs in context with other ferroelectric CDWs. Fig. \ref{fig:5} is a scatter plot of room-temperature zero gate bias resistance versus polarization which compares the CDW-FETs, labeled as H-H \#1-3 (red stars), to the resistances of different ferroelectric CDWs reported in the literature.\cite{mcquaid2017, choi2010, oh2015, mundy2017, wu2012, lindgren2017, sluka2012, jiang2020, werner2017, stolichnov2015, risch2022, Farokhipoor2011, seidel2009, zhang2023, Maksymovych2011, Liu2021}. The published reports are all for out-of-plane CDWs, with different materials shown in different colors and different shapes representing results obtained from different studies. The resistances obtained from the literature are either measured with two-point measurements on a conductive bottom substrate via probe-based techniques (denoted as solid markers) or by patterned top electrodes (hollow markers). The best comparison to our results are the patterned electrodes because they have a smaller channel to contact length ratio (0.3-1 versus 1.2) compared to the CDW-FETs measured in this study. We plot with respect to the polarization because CDW resistance is proportional to the amount of polarization bound charge at the interface.\cite{bednyakov2018} Despite having the lowest polarization, the CDW-FETs show resistance that ranges from on par to 2 orders of magnitude lower than the best reported values among all out-of-plane charged domain walls. Notably, H-H \#2 shows a minimum value of 3.1 k$\Omega$. The H-H \#1-3 show around 4 orders of magnitude difference in resistance. This device-to-device variation in resistance arises partly from the device geometry, but primarily due to variations in CDW disorders. We also observe the presence of Schottky barrier at the contact in H-H \#3 (Supplementary Fig. 6). The exact nature of the disorder depends on many uncontrolled parameters, including the interfacial twist angle, topography roughness, mechanical strain during stacking, and the exact phase of the material at the interface, and contributes to variation in resistance between devices.

To put these results into perspective, the small resistances in these CDW-FETs overcome a key limitation of most ferroelectric CDWs for high speed read-out circuits, where the domain wall currents typically fall short of the 0.1 $\mu$A required for 10 ns read times.\cite{sharma2022, Liu2021}


In summary, we demonstrate a robust method for generating highly conducting ferroelectric CDWs and integrating them into functional devices. This method of creating artificial CDWs should be universally applicable to any van der Waals ferroelectric with out-of-plane polarization. The artificial CDWs show two modes of conduction: high temperature trap assisted transport and low temperature VRH transport. Despite having the lowest polarization, these CDWs show the lowest resistance compared to the other ferroelectric CDWs reported in the literature. We also note that the conductivity of these devices is limited by disorder, and further improvement is possible by controlling and understanding the role of twist angle, interfacial roughness, and defects. Finally, similar artificial stacking of multiple polar domains and independently controlling each one should enable multiple conducting states, useful for multi-level data storage and memristor applications.



\section*{Methods}\label{Methods}
\subsection*{Fabrication of H-H CDWs}

We prepare the artificial H-H CDWs by stacking two \InSe{} flakes on top of each another. We first mechanically exfoliate two flakes on SiO\textsubscript{2} and polydimethyl-siloxane (PDMS) substrates and subsequently transfer the flake on PDMS on top of another using a standard dry-transfer technique.\cite{castellanos2014,nahid2024} All the exfoliation and transfer processes are performed inside a nitrogen glovebox. After stacking, we use PFM to confirm the polarization direction of each flake to identify the H-H CDWs.

\subsection*{Fabrication of Field-Effect Transistors}

We fabricate the \InSe{} single domain, H-T heterostructure, and H-H CDWs based FETs by standard electron-beam lithography, followed by metal deposition and lift-off. We spin-coat 495 A4 Poly(methyl methacrylate) (PMMA) at 3000 rpm for 60 s, followed by baking at 110$^{\circ}$ C for 20 minutes. We then spin-coat the second layer PMMA 950 A2 at 3000 rpm for 60 s and bake at 110$^{\circ}$ C for 25 minutes. Next, we write the electrode patterns using the Raith EBPG5150 Electron Beam Lithography System (current 2 nA, aperture 300 nm, dose 300 C/cm\textsuperscript{2}). The design of the electrodes ensure edge contact with the CDW layer. We develop the exposed PMMA by using a cold IPA and DI water mixture (weight ratio 3:1) and deposit 5 nm Ti, 60 nm Au, 10 nm Pd, and 60 nm Au using ebeam evaporation. Finally, we lift-off and subsequently clean the sample in acetone bath.

\subsection*{Electrical and Magneto-Transport Measurements}

We perform the electronic transport and magneto-transport measurements in Physical Property Measurement System (PPMS). We measure the output and transfer characteristics of the FETs using a Keithley 4200A-SCS Parameter Analyzer. For four-terminal magneto-transport measurements, we use SR830 lock-in amplifier and measure the voltage drop across the channel under a constant current bias of 10 nA and at 13 Hz frequency.

\subsection*{STEM Sample Preparation and Imaging}

Samples are first prepared for cross-sectioning by applying a protective coating of amorphous carbon via thermal evaporation. STEM lamella are then prepared via standard lift-out and thinning procedures in a Thermo Fisher Scientific Helios 600i DualBeam FIB-SEM with a cryo-can used during thinning. 

STEM imaging and EELS of cross-sectional samples were completed in a Thermo Fisher Scientific Themis Z aberration-corrected STEM operating at 300 kV with a convergence angle of 25.2 mrad.

\backmatter

\section*{Supplementary Information}\label{Supplementary Information}

\section*{Acknowledgments}

This work was entirely supported by NSF-MRSEC under Award Number DMR-2309037. This work was carried out in part at the Materials Research Laboratory Central Facilities at the University of Illinois. The authors acknowledge the use of facilities and instrumentation supported by NSF through the University of Illinois Materials Research Science and Engineering Center DMR-2309037.

\section*{Author Contributions}
Under A.M.v.d.Z's and S.W.N's supervision, S.M.N. fabricated the heterostructures and performed AFM and PFM characterizations. Under A.M.v.d.Z's and N.M's supervision, S.M.N. and H.D. fabricated and characterized the electrical transport and magneto-transport of the devices. Under A.S.'s and P.H.'s supervision, G.N. performed the STEM imaging. All authors read and contributed to the manuscript.

\section*{Data Availability}
 The data can be accessed for review at the Illinois Data Bank, DOI: 

\section*{Declarations}

The authors declare no competing financial interest.

\bibliography{sn-bibliography}

\end{document}